# MODEL DRIVEN ENGINEERING FOR SCIENCE GATEWAYS

David Manset[1], Richard McClatchey[2] and Hervé Verjus[3]

[1]GNUBILA France, Biomedical Applications, Argonay, France
[2]University of the West of England, CCCS, Bristol, UK
[3]University of Savoie, LISTIC, LS-LSE, Annecy-Le-Vieux, France
dmanset@gnubila.fr, richard.mcclatchey@uwe.ac.uk, herve.verjus@univ-savoie.fr

Keywords: MDE, SOA, ADL, Architecture-centric, Grid, Cloud, Science Gateway, Biomedical Research

Abstract: From n-Tier client/server applications, to more complex academic Grids, or even the most recent and promising industrial Clouds, the last decade has witnessed significant developments in distributed computing. In spite of this conceptual heterogeneity, Service-Oriented Architectures (SOA) seem to have emerged as the common underlying abstraction paradigm. Suitable access to data and applications resident in SOAs via so-called 'Science Gateways' has thus become a pressing need in various fields of science, in order to realize the benefits of Grid and Cloud infrastructures. In this context, authors have consolidated work from three complementary experiences in European projects, which have developed and deployed large-scale production quality infrastructures as Science Gateways to support research in breast cancer, paediatric diseases and neurodegenerative pathologies respectively. In analysing the requirements from these biomedical applications the authors were able to elaborate on commonly faced Grid development issues, while proposing an adaptable and extensible engineering framework for Science Gateways. This paper thus proposes the application of an architecture-centric Model-Driven Engineering (MDE) approach to service-oriented developments, making it possible to define Science Gateways that satisfy quality of service requirements, execution platform and distribution criteria at design time. An novel investigation is presented on the applicability of the resulting grid MDE (gMDE) to specific examples, and conclusions are drawn on the benefits of this approach and its possible application to other areas, in particular that of Distributed Computing Infrastructures (DCI) interoperability.

## 1 INTRODUCTION

Primarily developed by and for High Energy Physics (HEP), the Grid has been realised since the late 1990s as the next generation of information and communication technologies, after the Internet. Grid computing [1] promises to resolve many of the difficulties in facilitating massive data analyses to allow communities of end-users to collaborate without having to co-locate. Intrinsically distributed and highly heterogeneous, the Grid is the next logical step following the developments in high performance, high throuput and supercomputing.

The Grid is the product of collaborative developments worldwide. It often materializes as a set functions arranged in a so-called "middleware", i.e. a stack of commodity software sitting in and mediating between compute resources and user applications. Grid middleware are made of various types of services from low-level physical resources management, to computing power and storage capacity sharing, to more advanced information system and application scheduling services. Thus described, Grids are mostly implemented as Service Oriented Architectures (SOA) [2]. Given their functional scope and nature, Grids thus result in complex stratifications of software difficult to reuse, evolve and maintain [3]. Consequently, not only is the development of Grid-based applications a time-consuming, error prone and expensive task, but also are the resulting applications often hard-coded for specific configurations, technological platforms and physical infrastructures. The infrastructural functions offered by the Grid therefore need adaptation. This is what led research communities utilizing it to develop the concept of "Science Gateways".

Science Gateways represent an important emerging paradigm for providing integrated infrastructures. According to [4], a Science Gateway is a community-developed set of tools, applications, and data that are integrated via a portal or a suite of applications, usually in a graphical user interface, that is further customized to meet the needs of a specific community. Gateways enable users to access computing resources through a common and user-friendly interface.

However, given the underlying distributed computing infrastructures complexity, Science Gateways reuse and evolution is increasingly complex and the use of most classical engineering practices reveals inappropriate as few exhibit the necessary level of interoperability and flexibility required to import, integrate and to pass on the cumulated design data, information and knowledge to next generations [5]. There however exist engineering techniques such as architecture-centric design [6] which could help managing accidental difficulties faced with bridging conceptual gaps from abstraction to implementation and better adapting developments to evolving environments, such as Grids. Additionally, Model-Driven Engineering (MDE) [7] could help addressing models heterogeneity, separation of concerns, integration and interoperability.

The remainder of this paper thus attempts to characterize the specificities of Grid-based Science Gateway developments from practical examples in biomedical sciences. Section 2 reports on experiences carried out in three conceptually complementary infrastructures that address a broad spectrum of biomedical research requirements. Section 3 identifies common design issues faced in Science Gateways development, which section 4 then addresses by introducing a new MDE approach. The paper finally concludes on the significance of this research work and indicates experiments that could elaborate on new potential areas of application.

## 2 SCIENCE GATEWAYS IN BIOMEDICAL RESEARCH

With its roots grounded in HEP, the Grid required significant adaptation to be brought into and to serve the biomedical environment. The following sections report on three incremental Grid-based Science Gateways development experiences.

### 2.1 Breast Cancer, The EU FP5 MammoGrid Project

MammoGrid [8] aimed at utilizing the Grid as a digital repository to federate mammographic images and medical data, thereby allowing clinical researchers to store, share anonymously and analyze sensitive information acquired from various hospitals across Europe, in the context of specialized breast cancer studies. By doing so, MammoGrid made it possible for the first time to accumulate rare data samples into a common, secure and distributed repository needed to validate new breast cancer Computer Aided Detection (CAD) algorithms using the Standard Mammogram Format or SMF [11], while testing the actual feasibility and overall impact of providing automated radiographer second opinion in the cancer screening practice.

Developed between 2002 and 2005, MammoGrid adopted and adapted the first official release of the gLite Grid middleware [12], being issued by the Enabling the Grid for E-sciencE (EGEE) European project. At that time, the Grid resembled a Unix-like operating system managing distributed computing resources over a network, using specific command line interfaces. As it was the implementation of a new paradigm in computing carried out by large and geographically distributed communities, the form of the Grid used in MammoGrid was a rather complex, slow and heterogeneous software stack, difficult to install, configure and maintain. It was also not functional for instantaneous user interaction and was not regarded as sufficiently user-friendly by the biomedical research community. Biomedical researchers were thus hesitant in using it, as reported in [13]. Despite this, MammoGrid demonstrated for the first time the relevance of using this technology to support large-scale and automated second opinion and to allow clinical researchers to federate meaningful data into one shared environment.

### 2.2 Paediatric Diseases, The EU FP6 Health-e-Child Project

Elaborating on the MammoGrid model, the Health-e-Child project [14] then diversified Grid usage for biomedicine, by developing Decision Support Systems (DSS) and Knowledge Discovery tools supporting paediatricians in their daily work with integrated data in cardiology, especially in cardiomyopathies follow-ups, in rheumatology with juvenile arthritis diagnosis and in neuro-oncology with glioma evolution.

Health-e-Child was developed between 2006 and 2010 and it acknowledged the need for users to abstract from ongoing Grid developments in order to lower the barriers of adoption. Health-e-Child thus further developed the notion of a "Gateway" to the Grid, inserting a thin layer of abstraction services between the lower-level middleware and users, which would confine the unstable Grid under well-defined APIs. This thin Web services-based stack significantly improved the integration between new applications being developed in the project and the underlying Grid legacy. It also helped to convince non-IT users to adopt the technology, although performance remained an issue, as was reported in [15]. The Grid indeed remained too slow in manipulating data since it had been designed for long and non-fragmented runtimes, complex and highly versatile in nature. Deployed in five major hospitals across Europe and the USA, the solution however demonstrated significant reliability and security results.

### 2.3 Neuroimaging Biomarkers, The EU FP7 neuGRID Project

As a third generation infrastructure, the neuGRID project [17], attempted to further improve the Grid experience by pioneering a form of virtual laboratory for neuroscientists to develop, test and validate innovative new imaging biomarkers for neurodegenerative diseases. NeuGRID extended the idea of a "Science Gateway" to facilitate access to massive computing capacities.

NeuGRID was developed between 2008 and 2011 (and has since then has received further funding until 2015, under project name N4U). neuGRID based its architecture on the latest secure, reliable and performant Grid middleware products. It deployed a large-scale production quality infrastructure at specialized clinical centres, interconnected with the European Grid Initiative (EGI [16]), where it could access additional computing resources from. Although major improvements took place in the Grid, its evolving and heterogeneous nature encouraged neuGRID to further decouple its solution by adding new abstraction layers to form its Science Gateway. The latter relied on the following three pillars, as is further detailed in [17]: (1) Use of a so-called generic "gluing service" as part of the SOA to submit jobs to underlying Grids (see JavaGAT/SAGA [18] and neuGRID's gluing service [19] for more information). The gluing service abstracts upper layers of the system from the Grid specificities and is responsible for actual job submissions. (2) Use of a generic Web service wrapper in charge of on-the-fly orchestration and applying scheduling optimization techniques according to specified pipeline contents. (3) Instantiating a unique Web service wrapper per algorithm/pipeline to be published in the SOA, thus allowing (both atomic and composite) processing tasks to be discovered, composed and subsequently published in the system.

Each of these three substrates played a different but key role. While (1) introduced abstraction from Grids and thus allowed interacting with a wide variety of middleware, (2) took care of appropriately parameterizing (1), it also characterized commonalities of algorithms/ pipelines and opened a broad avenue to job scheduling optimization techniques (e.g. jobs grouping). Pillar (3), on the other hand, extended the parameterizing of (2) and turned these virtualized neuro-utilities into a set of standard services.

## 3 DESIGN ISSUES IN GRID-BASED SCIENCE GATEWAYS

Experiences over the last decade, a subset of which was presented in the previous section, demonstrate that the Grid has evolved from a very complex, slow and heterogeneous stack, difficult to install, configure and maintain into what is now regarded as a secure, reliable and maintained software. However, the Grid remains complex, evolving and heterogeneous. This is why applications being developed on top of, or integrating the Grid may risk becoming unsustainable, may lack interoperability, may remain complicated and can thus induce reluctance in users to adopt them. This motivates the case for Grid-based biomedical Science Gateways, which moreover deal with potentially sensitive medical data, which places more specific design constraints onto Grid infrastructures, in particular in terms of:

(a) Privacy, when sharing information that potentially identifies individuals. For example genetic profiles carrying DNA, unstructured data such as diagnostic reports sometimes encompassing patient's name and more, Magnetic Resonance (MR) images of patient brains allowing 3D reconstruction of patient's face etc.,

(b) Security, when sharing and storing data that potentially identifies individuals. Identifying data may be voluntarily shared for the sake of running for instance a clinical trial needing information on patients' living places for solving a given epidemiological question,

(c) Reliability, when storing and accessing medical data or clinical applications. Assisting physicians with decision support applications at the point of care may require highly available services in the infrastructure,

(d) Sustainability, when storing medical data as this can imply in some countries the ability to retrieve and make data accessible for 15 years or more.

In addressing the findings from [8], [10], [15] and [17], the authors assert the hypothesis that Grid-based biomedical Science Gateways should be designed as (1) Service Oriented Architectures (SOA), which (2) have specific Quality of Services (QoS) requirements, and (3) can be built on several technological platforms and physical resources. This is what Figure 1 illustrates. Such SOA-based, QoS-specific and multi-platform Science Gateways, are made of services exhibiting particular functions and properties in order to hide the Grid complexity and to help address community-specific issues like (a), (b), (c) and (d), formerly introduced.

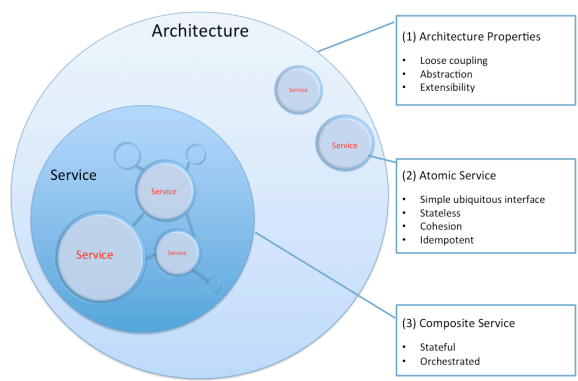

Figure 1. Science Gateway Architectural Style

Science Gateways enable the decoupling of new applications from evolving Grids, facilitate integration and transition to it, promote better reuse of software artefacts, and thereby potentially lower the barriers of user adoption. Figure 1 summarizes the basic architectural properties, which were unveiled thus far. Indeed, starting from the architecture level, i.e. (1), Science Gateways should follow the SOA style, in promoting abstraction, loose coupling and extensibility. Science Gateways should encompass component services, which can be specialized to target platforms, standards and technologies. Inner Science Gateway atomic services, i.e. wrapping low-level functions (2) should exhibit simple ubiquitous interfaces, be stateless, group coherent sets of functions and be idempotent. Composite services (3) on the other hand, (i.e. wrapping processes calling other services), should be stateful, so to store persistently important execution state information, and moreover be orchestrated. Science Gateways should therefore encompass mechanisms allowing the publication, discovery and composition of integrated services.

### 3.1 Science Gateways Engineering

Science Gateways should be parameterized/optimized according to non-functional requirements, such as, for instance, the expected level of reliability, security and privacy (i.e. QoS). Component services as identified in the former sections should therefore be assigned with QoS descriptive information accordingly at design time and the latter be mapped to architectural solutions, to be satisfied at runtime. Science Gateway architectures should be reusable, adaptable and portable to different research groups, execution platforms, technologies and physical infrastructures. Moreover, the deployment of such architectures may require taking into account distribution aspects, especially when under privacy, security, performance and/or reliability constraints. Thus, gateway architectures, properties and associated QoS, should be specified independently of any execution platforms, computing paradigms and programming languages.

### 3.2 Science Gateways Synthesis

From the MammoGrid, Health-e-Child and neuGRID experiences, the unveiled characteristics of Science Gateways indicate that a meta-model describing their architectural commonalities and properties could be designed, thereby allowing their reuse, adaptation and specialization to different fields of science. Science Gateways would thus significantly benefit from platform independence and their engineering should promote:

i. A high-level of abstraction, guaranteeing the Science Gateway model independence from any platform specificities,
ii. Models reuse, allowing the creation and use of basic building blocs,
iii. QoS properties specification, translating various types of non-functional requirements into design properties,
iv. Multi-platform portability, making it possible to port Science Gateways to different environments and technologies and
v. Distribution strategy formulation, enabling Science Gateways to have optimized deployments over target infrastructures and QoS.

### 4 LITERATURE REVIEW

In current research infrastructures, where utilizing the Grid implies its further adaptation, SOAs seem to have become the common abstraction paradigm to simplify access and developments, even though different standards and technologies may be applied across research projects and groups. SOA-based Science Gateways are thus emerging in various research fields and biomedical specialties, which operate most of the time for fixed QoS and execution platforms and are deployed over predefined physical infrastructures. Some offer customized Web-portals [20], thus simplifying access to the Grid infrastructure. Others focus more on scientific workflows [21], making the assumption that the infrastructure provides a sufficiently user-friendly access through which user applications can be designed as workflows. For the most advanced Science Gateways, a development framework [22] is provided, which allows developers to create and personalize new ones to their own needs ranging from the security model, to the privacy level, its reliability, the concrete Grid infrastructure to interface with, or even to the actual user interfaces.

The following synopsis table, Table 1, recalls the main criteria, as were identified in the former synthesis section, and which Science Gateway engineering approaches shall satisfy. This table allows comparing available approaches, while understanding their underlying concepts. In Table 1 references to the analysed approaches are provided in the left column, followed by a few keywords on their foundational paradigms and the five main comparison criteria.

| Ref. | Paradigm | Modeling Abstraction | Models Reuse | QoS Specification | Platform Portability | Distribution Strategies |
|---|---|---|---|---|---|---|
| [21] | Web portal and Workflow oriented | Yes* | Yes | No | No | Yes** |
| [22] | Web portal, SOA and workflow oriented | Yes* | No | Yes* | Yes* | Yes |
| [33] | Web portal and Workflow oriented | Yes* | Yes | No | No | Yes** |
| [34] | Web portal and Workflow oriented | Yes* | Yes | No | No | Yes** |
| [20] | Web portal and service | No | No | No | No | No |
| [35] | Web portal, ROA, and ORM oriented | Yes* | No | No | Yes | Yes** |
| [36] | Toolkit and SOA oriented | Yes* | No | No | No | Yes** |
| [37] | Web portal and service | No | No | No | No | No |

\* Only partially achieved.
\*\* Only made possible thanks to the workflow orientation.

Table 1. Literature Review in Science Gateways Engineering Approaches

Several conclusions can be drawn from this comparison. Firstly, the literature review demonstrates that simple service-based approaches do not address the identified criteria. Indeed, these approaches mainly facilitate the development of user interfaces by hiding the complexity of the underlying Grid, while they remain highly specific to the targeted technologies. On the other hand, Workflow-oriented solutions do exhibit interesting characteristics since they introduce abstraction and reuse of application models. They are consequently close to satisfying the identified requirements, although there is no approach yet tackling models reuse and quality of services at the same time. Finally, it is worth noting that approaches leveraging on abstraction, loose coupling and extensibility, i.e. utilizing SOAs, are the ones addressing best the Science Gateways engineering needs.

Given the lack of engineering methods available to address the identified criteria in a single and unified design process, the authors have been looking for candidate engineering techniques and their possible application. In particular, the proposed work has been motivated by the research carried out in SOA engineering and more specifically in architecture-based software developments [23]. Given that Science Gateways are sets of interconnected component services, architecture-centric software-based development applies particularly well since it allows the definition of distributed systems in terms of groups of components at a high-level of abstraction guaranteeing platform independence, enabling models reuse and, for some architecture-based approaches, expressing accompanying properties. Additionally, the authors considered the more recent Model Driven Engineering (MDE) [7] as a possible means to supplement architecture-based software development with a compositional technique to manage multi-platform complexity and thus automate adaptation/evolution. In the next section, readers will gain deeper understanding of the proposed combination of software engineering methods and be presented with the resulting "grid Model Driven Engineering" (gMDE) approach.

## 5 THE GRID MODEL DRIVEN ENGINEERING (GMDE)

### 5.1 gMDE Foundations

This paper introduces and tests a model-based engineering technique, which the authors propose to address the identified requirements in Science

Gateways engineering. The first ingredient used is a formal Architecture Description Language (ADL), the ArchWare Refinement Language (ARL) [24] to model and check Grid-based Science Gateways. Utilizing a formal architecture-centric method brings the necessary abstraction logic and mathematical foundation [25] to describe abstract software architectures, to model and test their architectural properties, and to ultimately transform these into concrete applications, i.e. the so-called process of refinement. The used formal Architecture-centric approach relies on languages and styles to describe applications, as well as tools for reasoning on architectural properties. It also introduces a development process that exploits and specializes iteratively abstract architecture descriptions into concrete applications, through stepwise refinement. This dimension of the proposed works is aimed to bring rigor and control into the Science Gateway engineering process. It addresses criteria (i) platform independence, and (ii) models reuse, while giving the foundations to express and check accompanying architectural properties (iii), such as QoS and target platforms. As the second ingredient, a Model-Driven Engineering (MDE) technique is proposed to promote models reuse and, thanks to the separation of concerns, to model transformations, to hide platform complexity and to refine abstractions by operating model transformations. MDE consequently supplements the design process with a compositional technique to manage complexity and to automate adaptation, utilizing a repository of "off-the-shelf" architectural constructs. It contributes to the proposed approach in improving flexibility and adaptability to changing environments, while allowing the long-term capitalization of architectural knowledge, thereby addressing the aspects of (iv) portability and (v) distribution in Science Gateways engineering.

Finally, a Domain Specific Language (DSL) [26] is introduced that allows modelling more specifically Grid-based Science Gateway architectures in terms of services and their interconnections. The DSL is encoded in the graphical user interface of the gMDE environment (gMDEnv), to facilitate the overall understanding and graphical design of Science Gateway solutions.

## 5.2 gMDE Design Process and Models

The grid Model Driven Engineering approach (gMDE) consists of a combination of existing and well-tested engineering techniques. In particular, gMDE builds on the work carried out by authors in the European FP5-funded ArchWare project [27], which developed a formal architecture-centric engineering toolkit of ADL [28] languages and accompanying toolkit. gMDE leverages on architecture-centric design to place the focus on coarse-grained system architecture specification, rather than coping up-front with implementation details. In doing so, software architects can design Science Gateways in terms of reusable and platform independent components (i.e. basic building blocs) and their interrelations. In paper [29], the authors introduced the foundational architecture-centric approach and toolset, which the novel gMDE engineering technique extends. Authors then presented the overall gMDE design process, with its eight models from the platform independent architecture specification (GEIM), to its specialization according to QoS (GECM) and platform (GETM) constraints, and finally to the (semi)-automatically generated source code (GESA) of the Science Gateway and its proposed distribution (GEDM) over the physical Grid infrastructure.

gMDE leverages on the model driven compositional dimension which it combines with architecture-centric refinement to translate non-functional concerns into architectural constructs, and then integrate them into the application model. A refinement step typically leads to a more detailed architectural model that increases the determinism of and preserves the properties associated with the abstract model. The ArchWare ARL language is the formal expression of these refinement operations [24]. ARL operates refinement operations by formally rewriting ARL architectural specifications using the Maude [25] formal rewriting logic.

## 6 APPLYING GMDE

The formerly introduced application areas are here explored successively in order to exemplify the application of the gMDE design process to solve identified engineering issues starting from a platform independent specification, and evolving to the concrete Science Gateway application. In order to simplify understanding, the given demonstration focuses on one stage of the design process per application area. Thus, a running example is taken from one end to the other.

### 6.1 Breast Cancer - Second Opinion

The MammoGrid Science Gateway encompasses a key set of commodity services. Firstly, authentication (Auth) and authorization (Authz)

```
gatewayArchitectureRef is style SOAScienceGateway
where {
    structure is {
        Portal is style serviceTypeRef where {
            structure is {… service internal structure
                description … }
            connection is { … service connections
                descriptions … }
            constraint is { … QoS and / or platform
                constraints mappings … }
        } …
        Auth is style serviceTypeRef where {
            structure is {… service internal structure
                description … }
            connection is { … service connections
                descriptions … }
            constraint is { … QoS and / or platform
                constraints mappings … }
        } …
        Authz is style serviceTypeRef where {
            structure is {… service internal structure
                description … }
            connection is { … service connections
                descriptions … }
            constraint is { … QoS and / or platform
                constraints mappings … }
        } …
        GridProxy is style serviceTypeRef where {
            structure is {… service internal structure
                description … }
            connection is { … service connections
                descriptions … }
            constraint is { … QoS and / or platform
                constraints mappings … }
        } …
        DataProxy is style serviceTypeRef where {
            structure is {… service internal structure
                description … }
            connection is { … service connections
                descriptions … }
```

Figure 3. MammoGrid Science Gateway Model

services, to login and access distributed resources uniformly, according to a security model derived from the requirements and that rules access rights and protects sensitive medical data. Secondly, a Portal service is offered to simplify access to complex workflows of underlying system functions, such as automated second opinion in the present case. Finally, a data staging service is included, which conforms to medical data standards (DICOM [30] and HL7 [31]), to enable users to upload data to the system for subsequent analyses. In MammoGrid, these legacy assets are kept independent of target back-ends (i.e. databases, Grid platform and

```
constraintName is constraintTypeRef {
    on a:architecture actions {
        actionRef elemRef is typeRef
            {… element description … }
    on b:architecturalElement actions {
        actionRef c .
        actionRef d
        …}}…
```

Figure 4. Constraint Meta-model

execution environments) and surrounding security thanks to abstraction services, hereinafter referred to as "Proxies" in the Science Gateway architecture.

In this context, the first use-case scenario focuses on the biomedical research Science Gateway model and its specialization to the quality of service needs of MammoGrid, in the light of offering a reliable automated second opinion service to physicians at the point of care. Figure 3 describes the MammoGrid platform independent Science Gateway model in the gMDE DSL formalism. The latter is automatically produced by the gMDENv interface (note that these descriptions are only partial extracts in order to simplify understanding). As can be noted, the gMDE DSL allows users to simply and quickly define a Science Gateway in terms of coarse-grained services. The gMDE DSL is the language used by the gMDEnv environment to assist and simplify the graphical creation of Science Gateway architectures and their specialization, until the concrete application source code can be produced. The gMDE

```
FT_reliability is qualityOfServiceProperty {
    on mammogridGateway:architecture actions {
        include FTConnector is connector {
            … connector architectural description …}
        on mammogridDataProxy
            :architecturalElement actions{
            replicate mammogridDataProxy to
                mammogridDataProxyClone0;
            unify
                mammogridDataProxy::ComsP0::Coms
                OutC0 with
                FTConnector::
                mammogridGridProxyComsP0::mammogrid
                GridProxyIncC0
            unify
                mammogridDataProxyClone0::ComsP0:
                :ComsOutC0 with
                FTConnector::
                mammogridGridProxyComsP0::
                mammogridGridProxyIncC0
}}…
```

Figure 5. QoS Architectural Pattern – GECM

```
behaviour is {
    archetype mammogridPortal is component {…} .
    archetype mammogridGridProxy is component{…}.
    archetype mammogridDataProxy is component {…}.
    archetype mammogridDataProxyClone0 is
component
    {…}.
    archetype FTConnector is connector {
        behaviour is {
            recursive value availabilityChecking is
                abstraction();
                {
                    if (serviceDown) value
                        serviceRedirectionURL :=
                        mammogridDataProxyClone0;
                        availabilityChecking();
                };
                compose { availabilityChecking() }
            } .
            recursive value readGridDBEntries is
                abstraction();   {…};
            recursive value clientDataRequest is
                abstraction();   {…}; ...
            compose {readGridDB() and
                clientDataRequest()}... }}} …
```

Figure 6. Refined Gateway Architecture - GEIM'

DSL allows users to describe Science Gateway architectural styles, for reuse "off-the-shelf", with predefined sets of components and accompanying requirements, and then to instantiate them as a new GEIM model. The GEIM is then translated into regular ARL for applying model transformations. Like the GEIM, the GECM and GETM constraint models reflecting QoS and target platforms are expressed in the gMDE DSL.

Figure 4 illustrates the meta-model of a non-functional constraint architectural construct. The

```
gLite3Proxy is executionPlatformProperty {
  on health-e-childGateway:architecture actions {
    on health-e-childGridProxy
      :architecturalElement actions{
        include gLiteGlueing is component {
          … component architectural description
        }
        unify
health-e-childGridProxy::ComsP0::ComsOutC0 with
  gLiteGlueing::ProxyComsP0::ProxyComsIncC0 .
        unify
health-e-childGridProxy::ComsP0::ComsInC0 with
  gLiteGlueing::ProxyComsP0::ProxyComsOutC0
}}…
```

Figure 7. Execution Platform Construct – GETM

latter describes how to redefine the concerned component(s) and its surroundings in order to solve the indicated requirement. This specification is in fact a simplified formalism for grouping relevant ARL refinement operations to be applied onto a given Science Gateway architecture to integrate the architectural construct. Once the GEIM model has been translated into ARL by gMDEnv, the first conceptual difference, which can be noted, is that the model no longer refers to services, but now manipulates components and connectors (i.e. the "C&C" style) onto which refinement operations can be applied.

From the quality of service constraint indicated in the GEIM model, here "--<reliability::level::3>--", the corresponding architectural construct is selected from the framework library. In the present case, the framework selects the "FT_Reliability" connector, as illustrated in Figure 5. This construct is then read by the framework and turned into lower-level ARL refinement operations, which are applied by rewriting logic onto the original GEIM model. The "FT_Reliability" construct is thus "weaved" in the Science Gateway architecture, resulting in the GEIM' description, reported in Figure 6, where the "mammogridDataProxy" service is replicated and made reliable with a load-balancing and fault-tolerant connector, acting as a switchtender to user requests. The construct thus applied, turns the automated second opinion application into a reliable service, supporting physicians in the screening process. In this first use-case scenario, a demonstration is given of how platform independent models (i) can be reused (ii), as well as how QoS constraints can be expressed and then solved by transformation (iii), thanks to the gMDE engineering technique, and using the gMDEnv framework.

## 6.2 Paediatric Cardiology – Similarity Search and Decision Support

In Health-e-Child, the Patient Browser interface allows physicians to run a similarity search over the entire database, along with customized clinical criteria to identify patients with similar conditions and access their treatments outcome. To do so, the Grid analyses all patient records throughout the connected databases and builds a similarity distance matrix based on the clinical weight attributed to discriminating medical variables. The result is sent back to the physician and displayed in specialized user interfaces, highlighting the patient population statistical distribution and potential clusters of identified similarities. In this second use-case, the objective is to adapt the Science Gateway architecture to a specific Grid middleware, making it possible to migrate existing Health-e-Child applications to the latest version of the Grid, without reengineering. Thus, starting from the Health-e-Child GEIM model, the execution platform constraint specified by the architect is extracted, i.e. "archetype health-e-childGridProxy is component {--<gridBackend::gLite::3.0>--" and the corresponding construct picked from the library, see Figure 7. Again, the construct is weaved into the GEIM Science Gateway architecture by transformation, resulting in a more specific GESM model. Thus, the "health-e-childGridProxy" architectural element is refined into a gLite v3.0 proxy, by integrating the "gLite3Proxy" component and connecting it to other existing elements' ports and connections as is dictated by the construct. Here, criteria (iv) multi-platform portability is partly demonstrated with adaptation of the Science Gateway to multiple Grids, thanks to the integration of platform specific constructs by successive refinement operations.

## 6.3 Neurodegenerative Disease - Disease Markers Validation

In neuGRID, neuroscientists can select datasets and specify new research hypotheses under the form of scientific workflows. Workflows are translated into a series of finer-grained tasks, which are sent for processing in the Grid. The latter orchestrates the workflow until its completion. The resulting outputs are stored in the Grid and pointers are sent back to the users. In this last scenario, authors assume that the neuGRID platform specific Science Gateway GESM model is finalised.

Thus entering the last stage of the gMDE design process, the GESM specification is turned into concrete source code by a mapping translation. This is achieved by specific parsers, which were developed to map the ARL concepts to different execution environments and programming languages. The translation is operated by a dedicated service in the gMDEnv framework. In the present case, the parsing granularity level is set to "Complex Objects", which indicates that first order components of the architecture are to be translated into software services, whereas subsequent order components correspond to simpler programming objects. In neuGRID, the targeted environment is the Globus 4.0 software. Thus, the GEMM parser produces corresponding service classes and accompanying Web services descriptors for deployment. The Science Gateway GESA source code is thus generated according to the target execution environment, to be further compiled and deployed. Compilation and deployment finally takes place thanks to the Grabber service, of the gMDEnv framework. The latter utilizes an ARL representation of the physical infrastructure (i.e. the GERM model) to understand its distribution and to deploy the Science Gateway according to what the architect has specified in the GEDM deployment model.

In this concluding use-case scenario, criteria (iv) multi-platform portability is demonstrated with Science Gateway code generation according to target execution environment, and (v) distribution is addressed (but not demonstrated) utilizing the GERM infrastructure representation.

## 7 CONCLUSIONS

The research work reported in this paper demonstrates the formulated approach to engineering Science Gateways. It showed from experimentation the feasibility of combining two existing and complementary engineering techniques towards the creation of gMDE [29]. Since this approach is based on the concepts of re-use and execution platform independence, the engineering framework is not limited to the Grid-based biomedical research domain. Indeed, the same approach can tackle other SOA-based developments. Thus, the benefits of using the gMDE are substantial. Formal application models designed under the presented framework are persistent and re-usable. One can use libraries of previously stored models (as templates) to design new applications. Furthermore the approach is scalable; one can extend the scope of the framework by providing new constraint and mapping models. Application of the presented technique is being foreseen in the area of self-adaptive systems, in particular on how computational applications can benefit from autonomic computing concepts and where (g)MDE can be used to impact on running architectures to reconfigure by themselves. In [32], self-adaptive capabilities were introduced in the Grid middleware itself, regardless of executed applications, in order to make it self-reconfigurable to QoS failure scenarios.

An interesting area of future research is the development of Cloud deployment strategies, based on step (4) of the gMDE process, in particular utilizing the GEDM deployment model. Indeed, similar to what was done with GridProxy services to abstract from Grid middleware specificities, Cloud Proxies could be defined as architectural design constructs and the QoS attributes turned into concrete deployment strategies brokering towards different Cloud (IaaS and PaaS) providers.